\documentclass[conference]{IEEEtran}

\usepackage{ifpdf}
\usepackage{cite}
\ifCLASSINFOpdf
\usepackage[pdftex]{graphicx}
\DeclareGraphicsExtensions{.pdf,.jpeg,.png}
\else
\usepackage[dvips]{graphicx}
\DeclareGraphicsExtensions{.eps}
\fi
\usepackage{graphicx}
\ifCLASSOPTIONcompsoc
\usepackage[caption=false, font=normalsize, labelfont=sf, textfont=sf]{subfig}
\else
\usepackage[caption=false, font=footnotesize]{subfig}
\fi
\usepackage{amsthm} 
\usepackage{balance}
\usepackage{epstopdf}
\usepackage{amsmath}
\usepackage{amssymb}
\usepackage{color}
\usepackage{balance}
\usepackage{array}
\usepackage{mathtools} 
\usepackage{algorithm,algorithmic} 
\usepackage{enumitem} 
\usepackage[mathcal]{euscript} 
\usepackage{tabularx} 
\usepackage{multirow} 
\usepackage{booktabs} 
\usepackage{makecell} 
\usepackage{url}

\graphicspath{{./Figures/}}

\newtheorem{Lemma}{Lemma}
\newtheorem{Theorem}{Theorem}
\newtheorem{Corollary}{Corollary}

\definecolor{blue}{rgb}{0.0, 0.28, 0.67}
\definecolor{red}{rgb}{0.59, 0.0, 0.09}
\definecolor{grey}{rgb}{0.52, 0.52, 0.51}

\newcommand{\src}{\mathrm{S}}
\newcommand{\des}{\mathrm{D}}
\newcommand{\ris}{\mathrm{R}}

\renewcommand{\vec}[1]{\boldsymbol{\mathrm{#1}}}

\newcommand{\Gsr}{G_{\src r}}
\newcommand{\Grd}{G_{r \des}}
\newcommand{\Lsr}{L_{\src r}}
\newcommand{\Lrd}{L_{r \des}}

\newcommand{\ids}{\mathrm{s}}
\newcommand{\idd}{\mathrm{d}}
\newcommand{\idg}{G}
\newcommand{\idl}{L}

\begin{document}

\title{Aerial Reconfigurable Intelligent Surface-Aided Wireless Communication Systems}

\author{\IEEEauthorblockN{
			Tri~Nhu~Do\IEEEauthorrefmark{1},
			Georges~Kaddoum\IEEEauthorrefmark{1}\IEEEauthorrefmark{2},
			Thanh~Luan~Nguyen\IEEEauthorrefmark{3},
			Daniel~Benevides~da~Costa\IEEEauthorrefmark{4},
			and~Zygmunt~J.~Haas\IEEEauthorrefmark{5}}
		\IEEEauthorblockA{\IEEEauthorrefmark{1}Department of Electrical Engineering, \'{E}TS, University of Qu\'{e}bec, Montreal, QC, Canada}
		\IEEEauthorblockA{\IEEEauthorrefmark{3}Faculty of Electronics Technology, Industrial University of Ho Chi Minh City (IUH), HCMC, Vietnam}
		\IEEEauthorblockA{\IEEEauthorrefmark{4}Future Technology Research Center, National Yunlin University of Science and Technology, Yunlin, Taiwan, R.O.C}
		\IEEEauthorblockA{\IEEEauthorrefmark{5}Computer Science Department, University of Texas at Dallas, Richardson, TX, U.S.A}
		\IEEEauthorblockA{ Emails: 
			\IEEEauthorrefmark{1}tri-nhu.do.1@ens.etsmtl.ca,
			\IEEEauthorrefmark{2}georges.kaddoum@etsmtl.ca, \\
			\IEEEauthorrefmark{3}nguyenthanhluan@iuh.edu.vn, 
			\IEEEauthorrefmark{4}danielbcosta@ieee.org,
			\IEEEauthorrefmark{5}haas@utdallas.edu
			}
	}

\maketitle

\begin{abstract}
In this paper\footnote{This paper was accepted to be presented at PIMRC'2021.}, we propose and investigate an aerial reconfigurable intelligent surface (aerial-RIS)-aided wireless communication system. Specifically, considering practical composite fading channels, we characterize the air-to-ground (A2G) links by Namkagami-m small-scale fading and inverse-Gamma large-scale shadowing. To investigate the delay-limited performance of the proposed system, we derive a tight approximate closed-form expression for the end-to-end outage probability (OP). Next, considering a mobile environment, where performance analysis is intractable, we rely on machine learning-based performance prediction to evaluate the performance of the mobile aerial-RIS-aided system. Specifically, taking into account the three-dimensional (3D) spatial movement of the aerial-RIS, we build a deep neural network (DNN) to accurately predict the OP. We show that: (\textit{i}) fading and shadowing conditions have strong impact on the OP, (\textit{ii}) as the number of reflecting elements increases, aerial-RIS achieves higher energy efficiency (EE), and (\textit{iii}) the aerial-RIS-aided system outperforms conventional relaying systems.
\end{abstract}

\begin{IEEEkeywords}
Reconfigurable intelligent surface, Nakagami-m, inverse-Gamma, outage probability, deep neural network
\end{IEEEkeywords}

\IEEEpeerreviewmaketitle

\section{Introduction}

Reconfigurable intelligent surfaces (RISs) have arisen as a promising solution for future sixth generation (6G) wireless communication systems \cite{DiRenzo_JSAC_2020,Do_arXiv_2021}.
An RIS is an array consisting of a large number of passive reflecting elements, in which each element can be programmed and electronically controlled to configure its phase-shift independently so that impinging waves can be reflected and steered toward an intentional direction.
Recently, in \cite{Basar_ACCESS_2019}, the authors specified the optimal phase-shift configuration for RIS-aided point-to-point communication systems. In \cite{Huang_TWC_2019}, the authors showed that RIS achieves more energy efficiency (EE) than the conventional amplify-and-forward (AF) relaying. In \cite{Bjornson_WCL_2020}, RIS was demonstrated to outperform the conventional decode-and-forward (DF) relaying in terms of EE. Very recently, in \cite{VanChien_CL_2020}, the authors provided various coverage and capacity analyses on RIS-aided dual-hop relaying systems.

Toward 6G, non-terrestrial communications have been recognized as sustainable and reliable tools for communicating over remote and hazardous areas. However, due to dynamics of spatial movements, there is a need of new fading models to accurately characterize both air-to-ground (A2G) downlink (DL) and uplink (UL) channels. In \cite{Sharma_CL_2020}, considering three-dimensional (3D) spatial movements, the outage performance of unmanned aerial vehicles (UAV)-aided dual-hop relaying was analyzed. Recently, in \cite{Bithas_TCOM_2020}, based on empirical data obtained in A2G trials, A2G channels were modeled using Nakagami-$m$ multipath fading and inverse-Gamma (IG) shadowing. In \cite{Bao_WCL_2020}, the authors used a deep neural network (DNN) to predict the secrecy performance of A2G communications. More recently, the concept of aerial RIS (aerial-RIS)-enabled communication systems was proposed in \cite{Lu_TWC_2021}; however, the authors modeled A2G links using the free-space path-loss model and neglected the fading and shadowing effects.  

\textit{Different from existing works}, in this paper, we propose a practical aerial-RIS-aided wireless communication system subject to composite fading channel model, where the small-scale fading follows a Nakagami-$m$ distribution whereas the large-scale shadowing follows an IG distribution. We analyze the outage performance of the aerial-RIS-aided system under such a practical fading channel model.
Next, we consider a mobile environment, in which the 3D spatial movement of aerial-RIS is modeled using the random waypoint mobility model (RWMM). With such a mobile system, locations of the aerial-RIS are treated as random variables (RVs), which makes the outage performance analysis intractable. Thus, relying on data-driven methods, we build a DNN that can be trained to predict the OP using pre-collected channel state information (CSI) and associated OP data.
The key contributions of the paper are summarized as follows:
\begin{itemize}
\item We propose \textit{a technical framework} to derive the OP of the proposed aerial-RIS-aided system. Specifically, in order to circumvent the appearance of parabolic cylinder functions, we use the \textit{moment-matching technique} to fit the distribution of the product of four different RVs to that of a $[\mathrm{Gamma/(Gamma^2)}]$ RV. The re-structured RV is transformed again to a mixture Gamma (MG) RV using Gaussian-Laguerre quadrature. We then use Laplace transform to derive the distribution of the sum of multiple transformed MG RVs. With this proposed technical framework, we obtain a tight approximate closed-form expression for the system OP.
\item Considering the 3D spatial movement of the aerial-RIS, whose mathematical performance analysis is intractable, we propose a detailed procedure to train and test the DNN, which is tailored to our proposed system. We show that the trained DNN accurately predicts the system OP. 
\item Representative results\footnote{The source code is published at \url{https://github.com/trinhudo/Aerial-RIS}} show that the OP is significantly sensitive to the considered composite fading model. With the DNN-based predicted OP results, we demonstrate the achievable EE of the aerial-RIS in terms of transmit power consumption under different reflecting element settings. Besides, we show that the aerial-RIS-aided system outperforms conventional cooperative relaying systems. 
\end{itemize}

\noindent \textbf{Notations}: $[\vec{v}]_r$ denotes the $r$-th element of vector $\vec{v}$, $\mathbb{E}(\cdot)$ denotes expectation operation, $\Gamma(\cdot)$ is the Gamma function \cite[(8.310.1)]{Gradshteyn2007}, $\Phi_2^{(K)} (\cdot)$ is the multivariate confluent hypergeometric function \cite[pp. 290]{Srivastava1985}, and $K_\nu(\cdot)$ is the modified Bessel function \cite[(8.407.1)]{Gradshteyn2007}.

\section{System and Chanel Models}

\begin{figure}[!t]
\centering
\includegraphics[width=.9\linewidth]{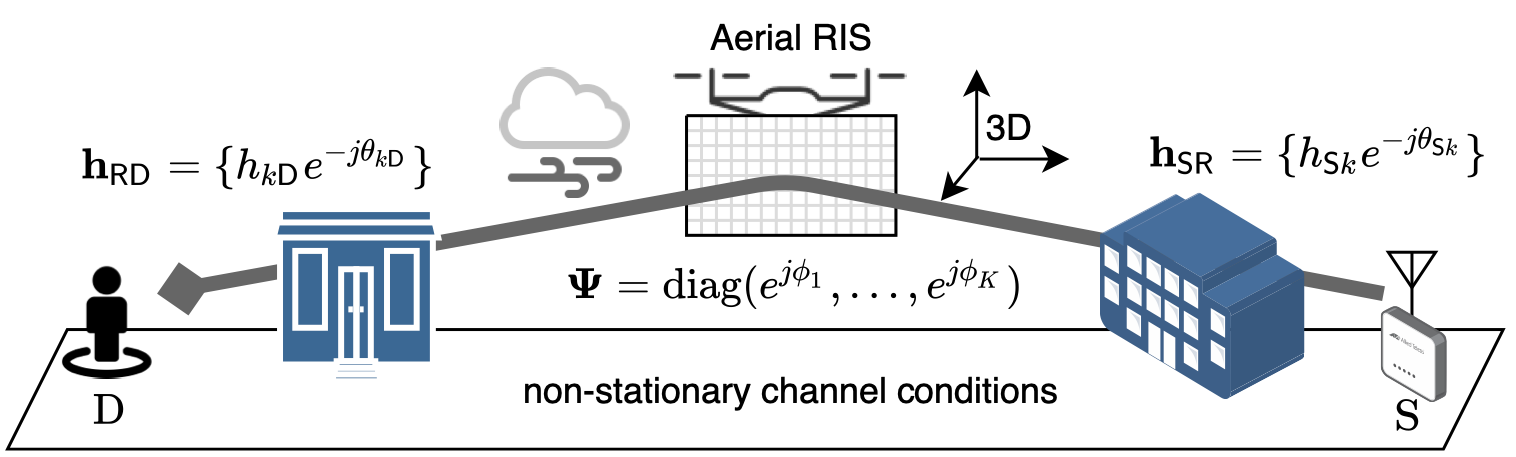}
\caption{Illustration of the aerial-RIS-aided wireless communication system.}
\label{fig_system}
\end{figure}

We consider a low complexity communication system that can be deployed for disaster relief. Considering that a source $\src$, e.g., a terrestrial transmitter, and a destination $\des$, e.g., a terrestrial receiver, are both equipped with a single-antenna, we assume that the $\src \to \des$ direct link is not available due to severe fading and shadowing caused by on-the-ground obstacles. Instead, the $\src \to \des$ communication is assisted by an aerial-RIS, $\ris$, which passively reflects signals from $\src$ to $\des$, as depicted in Fig.~\ref{fig_system}.

Suppose that the aerial-RIS has $N$ discrete reflecting elements, let $\vec{h}_{\src \ris} \in \mathbb{C}^{N \times 1}$ and $\vec{h}_{\ris \des} \in \mathbb{C}^{N \times 1}$ be the $\src\to\ris$ and $\des\to\ris$ \textit{complex channel coefficient vectors}, respectively. 
The properties of the aerial-RIS are characterized via the phase-shift matrix $\vec{\Psi} = \kappa \mathrm{diag}(e^{j \phi_1}, \ldots, e^{j \phi_N})$, where $\phi_r \in [0, 2\pi), r = 1,...,N,$ is the \textit{phase-shift} occurring at element $r$ of the aerial-RIS, and $\kappa \in (0,1]$ is the \textit{fixed amplitude reflection coefficient} \cite{Bjornson_WCL_2020}.
Let $s$, with $\mathbb{E}[|s|^2] = 1$, denote the transmit signal from $\src$. Thus, the received signal at $\des$ can be expressed as
\begin{align} \label{eq_received_signal}
	y = \sqrt{P_\src} \sum_{r=1}^N [\vec{h}_{\src \ris}]_r \kappa e^{j \phi_r} [\vec{h}_{\ris \des}]_r s + w_\des,
\end{align}
where $P_\src$ denotes the transmit power of $\src$ and $w_\des \sim \mathcal{CN}(0,\sigma^2)$ is the additive white Gaussian noise (AWGN) at $\des$ with zero mean and variance $\sigma^2$. 

Let $[\vec{h}_{\src\ris}]_r \triangleq \tilde{h}_{\src r}$ and $[\vec{h}_{\ris \des}]_r \triangleq \tilde{h}_{r \des}$. The polar representation of the \textit{complex channel coefficient} $\tilde{h}_\mathrm{c}$ can be expressed as $\tilde{h}_\mathrm{c} = h_\mathrm{c} e^{j \theta_\mathrm{c}}$, for $\mathrm{c} \in \{\src r, r \des\}$, where $h_\mathrm{c}$ is the \textit{magnitude}, i.e., $h_\mathrm{c} = \vert \tilde{h}_\mathrm{c} \vert$, and $\theta_\mathrm{c} \in [0, 2\pi)$ is the \textit{phase} of $\tilde{h}_\mathrm{c}$. Considering a practical composite fading channel, in which the small-scale fading, $G_\mathrm{c}$, is modeled as a Nakagami-$m$ RV and the large-scale shadowing, $L_\mathrm{c}$, is modeled as an IG RV, we have that $ h_\mathrm{c} = L_\mathrm{c} G_\mathrm{c}$. More specifically, the Nakagami-$m$ probability density function (PDF) of $G_\mathrm{c}$ is given by \cite{Matlab2021a}
\begin{align} \label{eq_PDF_Naka}
	f_{G_\mathrm{c}} (x ; m, \Omega) =  
	2 \left(\frac{m}{\Omega}\right)^m \frac{1}{\Gamma(m)} x^{(2m - 1)} e^{\frac{- m}{\Omega} x^2} , x>0,
\end{align}
where $m$ and $\Omega$ are the \textit{shape} and the \textit{spread} parameters  of the distribution, respectively. The inverse-Gamma PDF of $L_\mathrm{c}$ is given by \cite{Peebles2000}
\begin{align} \label{eq_PDF_IG}
	f_{L_\mathrm{c}} (x;\alpha, \beta) = \frac{\beta^\alpha}{\Gamma(\alpha)} x^{(-\alpha - 1)} e^{\frac{-\beta}{x}}, x>0,
\end{align}
where $\alpha > 1$ and $\beta$ are the \textit{shape} and the \textit{scale} parameters of the distribution, respectively. 

From \eqref{eq_received_signal}, the end-to-end (e2e) instantaneous achievable capacity [b/s/Hz] of the system can be expressed as
\begin{align}
	R = \max_{\phi_1, \ldots,\phi_N} \log_2 \bigg( 1 + \frac{P_\src}{\sigma^2} \bigg| \kappa \sum_{r=1}^N [\vec{h}_{\src \ris}]_r e^{j \phi_r} [\vec{h}_{\ris \des}]_r \bigg|^2 \bigg).
\end{align}

\section{Outage Performance Analysis}

The outage probability of the system is the probability that the instantaneous mutual information of the system falls below a pre-defined target spectral efficiency (SE), $R_\mathrm{th}$, [b/s/Hz], which can be mathematically expressed as
\begin{align} \label{eq_outage}
 \mathrm{P_{out}} &= \Pr(R < R_\mathrm{th}) \nonumber\\
  &= \Pr(\gamma < \gamma_\mathrm{th}),
\end{align}
where  $\gamma_\mathrm{th} \triangleq 2^{R_\mathrm{th}} - 1$, and $\gamma$ denotes the e2e receive signal-to-noise ratio (SNR) at $\des$, which is expressed as
\begin{align} \label{eq_snr}
	\gamma = \max_{\phi_1,\ldots,\phi_N} \bar{\gamma} \bigg| \kappa \sum_{r=1}^N L_{\src r} G_{\src r} e^{j (\phi_r + \theta_{\src r} + \theta_{r \des}) } G_{r \des} L_{r \des} \bigg|^2 ,
\end{align} 
where $\bar{\gamma} = P/\sigma^2$ denotes the average transmit SNR [dB]. 

In order to address \eqref{eq_outage}, we first setup the optimal configuration of the phase-shift matrix, i.e., $\vec{\Psi}^\star$, such that the end-to-end SNR in \eqref{eq_snr} is maximized. Following \cite{Basar_ACCESS_2019}, \cite{Bjornson_WCL_2020}, and \cite{VanChien_CL_2020}, for each reflecting element of the aerial-RIS, we configure $\phi_r^\star = - (\theta_{\src r} + \theta_{r \des}), \forall r$. Consequently, the OP in \eqref{eq_outage} can be re-expressed as
\begin{align} \label{eq_Psi}
\mathrm{P_{out}} = \Pr \bigg(\bar{\gamma} \kappa^2 \bigg| \sum_{r=1}^N L_{\src r} G_{\src r} G_{r \des} L_{r \des} \bigg|^2 < \gamma_\mathrm{th} \bigg).
\end{align}

First, let $W_r \triangleq \Lsr \Gsr \Lrd \Grd$, $G_r \triangleq \Gsr \Grd$, $L_r \triangleq \Lsr \Lrd$, and $\tilde{L}_r \triangleq 1 /\sqrt{L_r}$, thus, $W_r = G_r / \tilde{L}_r^2$. Next, we will show that the distribution of $W_r$ can be matched to a $[\mathrm{Gamma/(Gamma^2)}]$ distribution. To this end, we first propose a distribution matching for $G_r$ and $\tilde{L}_r$, as presented in the following Lemma.
Let $X$ be a Gamma RV with the PDF being given by \cite{Matlab2021a}
\begin{align}
f_{X} (x; \nu, \zeta) = \frac{1}{ \zeta^{\nu} \Gamma(\nu)}  x^{(\nu - 1)} e^{ \frac{-x}{\zeta} }, x>0,
\end{align}
where $\nu$ and $\zeta$ denote the \textit{shape} and the \textit{scale} parameters, we denote $X \sim \mathrm{Gamma} (\nu, \zeta)$.

\begin{Lemma} \label{lemma_matching_Gamma}
The distribution  of $G_r$ and $\tilde{L}_r$ can be approximately matched to a Gamma distribution as
\begin{align} 
	G_r &\overset{\mathrm{approx.}}{\sim} \mathrm{Gamma} (m_\idg, \Omega_\idg / m_\idg ), \label{eq_G_r_Gamma} \\
	\tilde{L}_r &\overset{\mathrm{approx.}}{\sim} \mathrm{Gamma} (m_\idl, \Omega_\idl / m_\idl), \label{eq_L_r_Gamma}
\end{align}
respectively, where
$\Omega_\idg \triangleq \frac{ \Gamma(m_\ids + 1/2) \Gamma(m_\idd + 1/2) }{\Gamma(m_\ids) \Gamma(m_\idd)} \Upsilon_\idg^{-1/2}$, where 
$\Upsilon_\idg \triangleq (m_\ids m_\idd)/(\Omega_\ids \Omega_\idd)$,
and 
$m_\idg \triangleq \frac{\Omega_\idg^2}{\Omega_\ids \Omega_\idd - \Omega_\idg^2}$;
and 
$\Omega_\idl \triangleq \Gamma(\alpha_\ids + 1/2) \Gamma(\alpha_\idd + 1/2) / [\sqrt{\beta_\ids \beta_\idd} \Gamma(\alpha_\ids) \Gamma(\alpha_\idd)]$ 
and 
$m_\idl \triangleq \Omega_\idl^2/ [(\alpha_\ids \alpha_\idd)/(\beta_\ids \beta_\idd) - \Omega_\idl^2]$.
\end{Lemma} 

\begin{IEEEproof}
The proof is provided in Appendix \ref{apx_proof_lemma_GG_LL}.
\end{IEEEproof}

To further the analysis, we present the following Theorem.
\begin{Theorem}\label{theorem_cdf_Z2}

Let $Z \triangleq \sum_{r=1}^N W_r = \sum_{r=1}^N \Gsr \Lsr \Grd \Lrd$, an approximate closed-form expression for the cumulative distribution function (CDF) of $|Z|^2 $ can be attained, as in \eqref{eq_cdf_Z2_a}, on the top of the next page, 
\begin{figure*}
\begin{align} \label{eq_cdf_Z2_a}
	F_{|Z|^2} (z) 
	\approx 
	\frac{1}{\Gamma(N m_\idg +1)} z^{\frac{N m_\idg}{2}} \sum_{\Xi_\tau, \forall \tau} \binom{N}{\tau_1,...,\tau_K} 
	\bigg[ \prod_{k=1}^K \xi_k^{v_k}\bigg] \Phi_2^{(K)} \bigg(m_\idg \tau_1, ..., m_\idg \tau_K ; 1+N m_\idg; -\frac{\sqrt{z}}{\zeta_1},..., - \frac{\sqrt{z}}{\zeta_K} \bigg),
\end{align} 
\hrule
\end{figure*}
where $\binom{N}{ \tau_1,\dots,\tau_K } \triangleq \frac{ N! }{ \tau_1 ! \dots \tau_K! }$, $\Xi_\tau \triangleq (\tau_1, ..., \tau_K)^{(\tau)}$ denotes a possible combination of $\tau_1, ..., \tau_K$, where $\tau_k \in \{\tau_1, ..., \tau_K\}$ are non-negative integers satisfying $\sum_{k=1}^K \tau_k = N$.
\end{Theorem}

\begin{IEEEproof}
Let $\Lambda_\idg \triangleq \Omega_\idg / m_\idg$ and $\Lambda_\idl \triangleq \Omega_\idl / m_\idl$. Invoking Lemma~\ref{lemma_matching_Gamma}, the approximate PDF of the $[\mathrm{Gamma/(Gamma^2)}]$ RV, i.e., $W_r = G_r/\tilde{L}_r^2$, is obtained as
\begin{align} \label{eq_pdf_W_r_a}
	f_{W_r} (y) \approx \frac{ \Lambda_\idl^{-m_\idl}}{\Gamma(m_\idl)} \int_0^\infty x^{m_\idl + 1} f_{G_r} (y x^2) e^{- \frac{x}{\Lambda_\idl}} d x.
\end{align}
In order to derive a tractable expression of the integral in \eqref{eq_pdf_W_r_a}, we rely on the Gaussian-Laguerre (G-L) quadrature \cite{Abramowitz1965}, which yields
\begin{align}
	f_{W_r} (y) \approx \sum_{k=1}^K \psi_k \frac{y^{m_\idg - 1}}{\Gamma(m_\idg)} e^{ - \frac{y}{\zeta_k}}, y>0,
\end{align}
where $\zeta_k = \frac{\Lambda_\idg}{  (\mathrm{z}_k \Lambda_\idl)^2}$, $\psi_k = \frac{\mathrm{w}_k}{\Gamma(m_\idl)} \frac{\mathrm{z}_k^{m_\idl - 1}}{\zeta_k^{m_\idg}}$, $\mathrm{w}_k$ and $\mathrm{z}_k$ are the weight factors and abscissas of the G-L quadrature, respectively \cite{Abramowitz1965}, and $K$ denotes the number of terms in the G-L quadrature. After normalization, i.e., $f_{W_r}(y) \leftarrow  f_{W_r} (y) / \int_0^\infty f_{W_r} (y) \mathrm{d}y$, an approximate closed-form expression for the PDF of $W_r$ can be attained as
\begin{align} \label{eq_pdf_W_r}
	f_{W_r} (y) \approx \sum_{k=1}^K \xi_k \frac{y^{m_\idg - 1}}{\Gamma(m_\idg)} e^{-\frac{y}{\zeta_k} },
\end{align} 
where $\xi_k = \psi_k / \big[ \sum_{i=1}^K \psi_i \zeta_i^{m_\idg} \big]$. From \eqref{eq_pdf_W_r}, it is apparent that $W_r$ follows a MG distribution.

We now turn our focus to $Z = \sum_{r=1}^N W_r$. First, the Laplace transform of $W_r$, $\mathcal{L}_{W_r} (v) \triangleq \mathbb{E}_{W_r} [e^{-v W_r}]$, can be obtained as
\begin{align}
	\mathcal{L}_{W_r} (v) = \int_0^\infty e^{-v y} f_{W_r} (y) d y = \sum_{k=1}^K \xi_k \left( \frac{1}{\zeta_k} + v\right)^{-m_\idg}.
\end{align}
By performing the Laplace transform for $Z$ and after some derivation steps, the CDF of $Z$ can be expressed as
\begin{align} \label{eq_CDF_Z_a}
	F_Z (z) = \mathcal{L}^{-1} \bigg\{ \frac{1}{v} \bigg[\sum_{k=1}^K \xi_k \bigg( \frac{1}{\zeta_k} + v \bigg)^{-m_\idg}\bigg]^N; v, z\bigg\},
\end{align}
where $\mathcal{L}^{-1}\{  H(v), v,z \}$ specifies the inverse Laplace transform of $H(v)$ from $v$-domain to $z$-domain. Invoking the multinominal theorem, \eqref{eq_CDF_Z_a} is re-expressed as
\begin{align}
	F_Z (z) &= \sum_{\Xi_\tau, \forall \tau} \binom{N}{\tau_1,...,\tau_K} \bigg[\prod_{k=1}^K \xi_k^{\tau_k}\bigg] \nonumber \\
	&\quad\times  \mathcal{L}^{-1} \bigg\{ \frac{1}{v} \prod_{k=1}^K \bigg( \frac{1}{\zeta_k} + v\bigg)^{-\tau_km_\idg}; v,z \bigg\}.
\end{align}
Given the linearity property of the Laplace transform, we have 
\begin{align}
	&\mathcal{L}^{-1} \bigg\{ \frac{1}{v} \prod_{k=1}^K \bigg( \frac{1}{\zeta_k} + v\bigg)^{-\tau_km_\idg}; v,z \bigg\} = \frac{1}{\Gamma(N m_\idg + 1)}\nonumber \\
	&\quad\times \mathcal{L}^{-1} \bigg\{ \frac{\Gamma(N m_\idg + 1)}{v^{N m_\idg + 1}} \!\prod_{k=1}^K \!\!\! \bigg( \frac{1}{\zeta_k v} + 1\bigg)^{-\tau_k m_\idg} \!\!\!\!\!\!\!\! ; v,z \bigg\},
\end{align}
and by making use of \cite[Eq. (10)]{Martinez_TCOM_2016}, after some mathematical manipulations/simplifications, an approximate closed-form expression for the CDF of $Z$ can be derived as
\begin{align} \label{eq_CDF_Z_d}
	&F_Z (z) \approx \frac{z^{N m_\idg}}{\Gamma(N m_\idg + 1)} \sum_{\Xi_\tau, \forall \tau} \binom{N}{\tau_1,...,\tau_K} \bigg[\prod_{k=1}^K \xi_k^{\tau_k}\bigg] \nonumber \\
	&\times \Phi_2^{(K)} \bigg(m_\idg \tau_1 ,..., m_\idg \tau_K; N m_\idg + 1; -\frac{z}{\zeta_1},...,-\frac{z}{\zeta_K} \bigg).
\end{align}
Since  $F_{X^2} (x) = F_X (\sqrt{x}), x>0$, one can attain the CDF of $|Z|^2$ as in \eqref{eq_cdf_Z2_a}. This completes the proof of Theorem \ref{theorem_cdf_Z2}.
\end{IEEEproof}

\begin{Corollary}
Invoking Theorem \ref{theorem_cdf_Z2}, an approximate closed-form expression for the system OP can be attained as 
\begin{align} \label{eq_OP_end}
\mathrm{P_{out}} \approx F_{|Z|^2} (\gamma_\mathrm{th}/ (\bar{\gamma} \kappa^2)).
\end{align}
\end{Corollary}

\section{DNN-based Performance Prediction}

\begin{figure}[t]
\centering
\includegraphics[width=.6\linewidth]{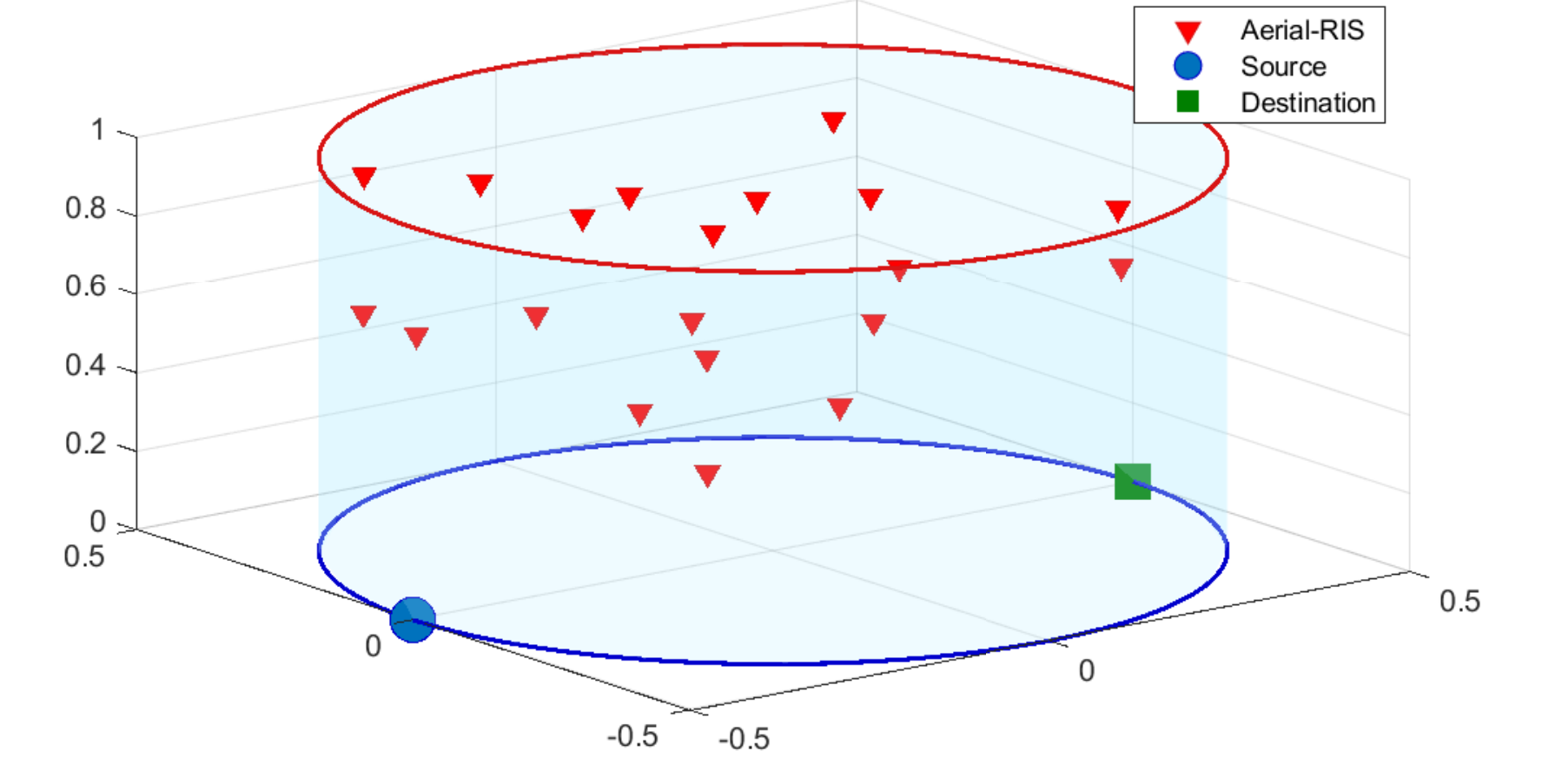}
\caption{Illustration of the 3D spatial movement of the aerial-RIS.}
\label{fig_3D_position}
\end{figure}

Given the system model considered in the previous sections, we further assume that the 3D spatial movement of the aerial-RIS is characterized by the RWMM, the samples of its locations can be drawn from a homogeneous 3D Poisson process. For the sake of exposition, we further assume that the aerial RSI spatially moves in a 3D cylinder, as depicted in Fig.~\ref{fig_3D_position}. Consequently, each element of the cylindrical coordinate of $\ris$ can be generated from a Uniform distribution, as shown in Table \ref{table_parameters}. It is noted that with the consideration of 3D movement, distances between nodes in the proposed system are now also RVs, which makes the derivation of the system OP defined in \eqref{eq_Psi} infeasible, mainly because of the multivariate confluent hypergeometric function. To overcome this hurdle, we treat the problem of finding the system OP as \textit{a regression problem in supervised learning}. In particular, we generate a data set that comprehensively characterizes the considered system. By being trained by such a data set, the developed DNN is capable of accurately predict the OP under various system settings.

\begin{table}[!h] 
\centering
\caption{Input Parameters for DNN Training and Testing}
\begin{tabularx}{\linewidth}{l X || l X}
\Xhline{2\arrayrulewidth}
\textbf{Inputs} & \textbf{Values} & \textbf{Inputs} & \textbf{Values} \\
\Xhline{2\arrayrulewidth}
$\bar{\gamma}$ [dB] & $[5-\epsilon_\gamma, 5+\epsilon_\gamma]$
& $N$ & $[20-\epsilon_N, 20+\epsilon_N]$ \\
$\omega_\ris$ & $\sim \mathcal{U}(0,2\pi)$  
& $\mathrm{r}_\ris$ &  $\sim 0.5 \sqrt{\mathcal{U} (0,1)}$\\
$m_\mathrm{c}$ & $[2-\epsilon_m, 2+\epsilon_m]$ 
& $\mathrm{h}_\ris$ & $\sim \mathcal{U} (0,1)$ \\
$\alpha_\mathrm{c}$ & $[2.5-\epsilon_\alpha, 2.5+\epsilon_\alpha]$ 
& $\beta_\mathrm{c}$ & $[1-\epsilon_\beta, 1+\epsilon_\beta]$ \\
$\eta$ & $[2.7-\epsilon_\eta, 2.7+\epsilon_\eta]$ & $R_\mathrm{th}$ & $[5-\epsilon_R, 5+\epsilon_R]$ \\
\Xhline{2\arrayrulewidth}
\end{tabularx}
\label{table_parameters}
\end{table}

\begin{figure}[t]
\centering
\includegraphics[width=.7\linewidth]{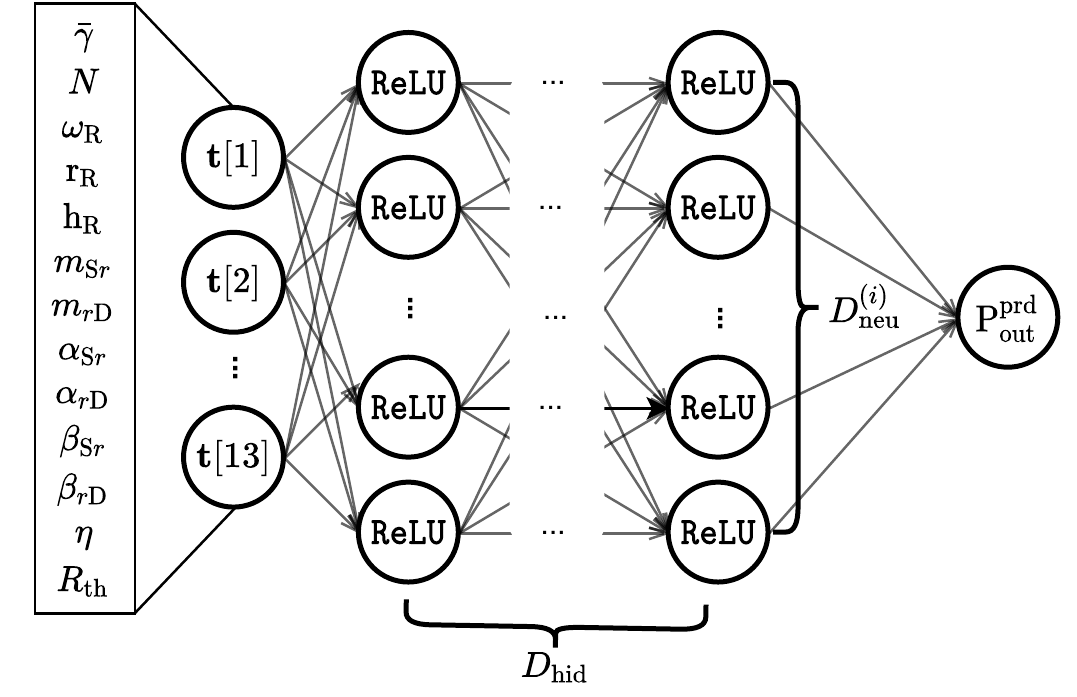}
\caption{Illustration of the design DNN for the considered regression problem.}
\label{fig_DNN}
\end{figure}

\subsection{3D Spatial Movement Modeling}
Let $\omega_\ris$, $\mathrm{r}_\ris$, and $\mathrm{h}_\ris$ respectively denote azimuth, radial distance, and height of the cylindrical coordinate of $\ris$. Without loss of generality, we consider a unit normalization cylinder, i.e., $\omega_\ris \in [0,2\pi]$, $\mathrm{r}_\ris \in [0,0.5]$, and $\mathrm{h}_\ris \in [0,1]$. Thus, the 3D Cartesian coordinates of $\ris$, $(x_\ris,y_\ris, z_\ris)$, are converted to $x_\ris = \mathrm{r}_\ris \sin \omega_\ris$, $y_\ris = \mathrm{r}_\ris \cos \omega_\ris$, and $z_\ris = \mathrm{h}_\ris$. Assume that the 3D Cartesian coordinates of $\src$ and $\des$ are $(-0.5,0,0)$ and $(0.5,0,0)$, respectively, a distance between two nodes is calculated as $d_\mathrm{AB} = \sqrt{(x_\mathrm{A} - x_\mathrm{B})^2 + (y_\mathrm{A} - y_\mathrm{B})^2 + (z_\mathrm{A} - z_\mathrm{B})^2}$, where $\{\mathrm{A,B}\} \in \{\src,\des,\ris\}$. 

\subsection{DNN Construction, Training, and Testing}

\subsubsection{The Structure of The DNN} Our developing DNN is a feed-forward neural network, consisting of one input layer, $D_\mathrm{hid}$ hidden layers, and one output layer, as depicted in Fig.~\ref{fig_DNN}. The input layer has $13$ neurons, corresponding to $13$ system parameters listed in Table~\ref{table_parameters}. Each hidden layer $i$, $i =1,...,D_\mathrm{hid}$, has $D_\mathrm{neu}^{(i)}$ neurons, and uses the rectified linear unit ($\mathrm{ReLU}$) function as activation function. The output layer has one neuron, and uses the linear function as its activation function to return the predicted OP, $\mathrm{P}_\mathrm{out}^\mathrm{prd}$.

\begin{algorithm}[t]
\renewcommand{\thealgorithm}{1}
	\caption{Procedure of training and testing the DNN}
	\begin{algorithmic}[1]
		\renewcommand{\algorithmicrequire}{\textbf{Input:}}
		\renewcommand{\algorithmicensure}{\textbf{Output:}}
		\REQUIRE Inputs parameters; DNN settings: $D_\mathrm{hid} =5$, $D_\mathrm{neu}^{(i)} =128$, $\mathrm{RMSE}_\mathrm{th}=2\times10^{-2}$, learning rate $\mathrm{lr} = 10^{-3}$
		\ENSURE  A trained DNN
		\STATE Draw $\mathcal{S}_\mathrm{trn}$, $\mathcal{S}_\mathrm{val}$, and $\mathcal{S}_\mathrm{tes}$ sets from the data set
		\STATE Create the DNN's structure using Keras and TensorFlow
		\WHILE {$(\mathrm{RMSE} \geq \mathrm{RMSE}_\mathrm{th})$}
		\STATE Dynamically adjust $D_\mathrm{hid}$, $D_\mathrm{neu}^{(i)}$, $\mathrm{l_r}$, number of epochs
		\STATE Use $\mathcal{S}_\mathrm{trn}$ and $\mathcal{S}_\mathrm{val}$ to train, validate, and then save the validated DNN as $\mathtt{validatedDNN.h5}$
		\STATE Feed $\mathcal{S}_\mathrm{test}$ into $\mathtt{validatedDNN.h5}$, obtain $\mathrm{RMSE}$
		\ENDWHILE
		\RETURN $\mathtt{trainedDNN.h5}$ 		
	\end{algorithmic} 
	\label{alg_DNN_train_test}
\end{algorithm}

\begin{figure} [!t] 
\centering
\includegraphics[width=.9\linewidth]{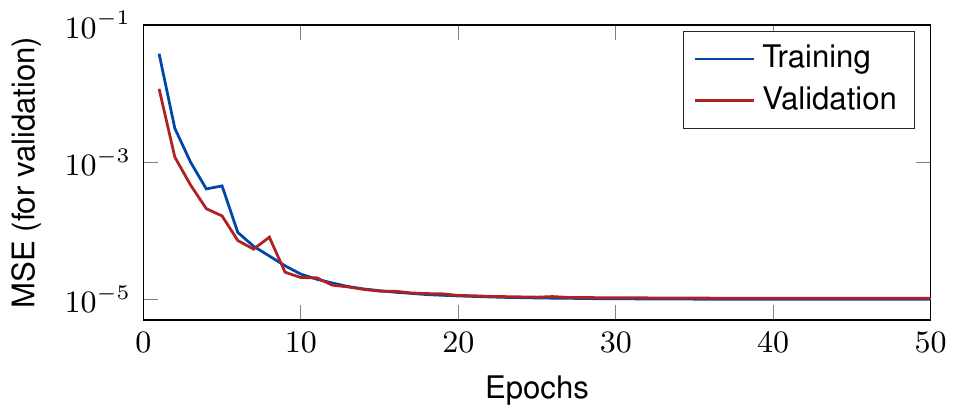}
\caption{The convergence of MSE in training and validating the DNN.}
\label{fig_validation}
\end{figure}

\subsubsection{Data Set}
Each sample $i$ of our data set $\mathcal{S}$ is a row vector, i.e., 
$\mathrm{Data}[i] = [\vec{t}[i],\mathrm{P}_{\mathrm{out},\textit{i}}^\mathrm{sim}]$, where $\vec{t}[i]$ is a feature vector including all input parameters listed in Table~\ref{table_parameters}. Each feature vector $\vec{t}[i]$ is used to create real-value CSI sets, which are fed into a  Monte-Carlo simulation, and returns a unique corresponding $\mathrm{P}_{\mathrm{out},\textit{i}}^\mathrm{sim}$. Totally, we create $10^5$ samples, i.e., $\mathrm{Data}[i], i= 1,...,10^5$, and concatenate them to create the data set.
We then divide the data set into training set, $\mathcal{S}_\mathrm{trn}$, validation set, $\mathcal{S}_\mathrm{val}$, and test set, $\mathcal{S}_\mathrm{tes}$, with ratio $80\%$, $10\%$, and $10\%$, respectively. 

Using the generated data sets, the DNN is trained and tested following Algorithm~\ref{alg_DNN_train_test}, where the mean squared error (MSE) is defined as $\mathrm{MSE} = \frac{1}{|\mathcal{S}_\mathrm{tes}|} \sum_{i=0}^{|\mathcal{S}_\mathrm{tes}|-1} \left(\mathrm{P}_\mathrm{out}^\mathrm{prd} - \mathrm{P}_\mathrm{out}^\mathrm{tes}\right)^2$, and the root MSE (RMSE) is defined as $\mathrm{RMSE} = \sqrt{\mathrm{MSE}}$.

\section{Numerical Results and Discussions}

In Nakagami-$m$ fading, the shape parameter $m$ indicates the fading severity and the scale parameter $\Omega$ takes into account the large-scale fading effect, i.e., $\Omega_\mathrm{c} = d_\mathrm{c}^{-\eta}$, for $\mathrm{c} \in \{\src r, r \des\}$, where $\eta$ denotes the path-loss exponent. 
In IG shadowing, $\alpha$ indicates the shadowing severity and $\beta$ is normalized with respect to $\bar{\gamma}$. Parameter settings are: $\epsilon_{\bar{\gamma}}=15$, $\epsilon_N  =10$, $\epsilon_m = \epsilon_\alpha = 0.5$, $\epsilon_\eta = 0.3$, $\epsilon_\beta = 0.2$, $\epsilon_R = 3$, $K=30$, and a total of $10^5$ sets of parameters are generated.

In Fig.~\ref{fig_validation}, we evaluate the accuracy of the training using the validation set. As can be seen, the MSE converges after $30$ epochs and is lower than $10^{-4}$.

\begin{figure} [!t]
\centering
\includegraphics[width=.9\linewidth]{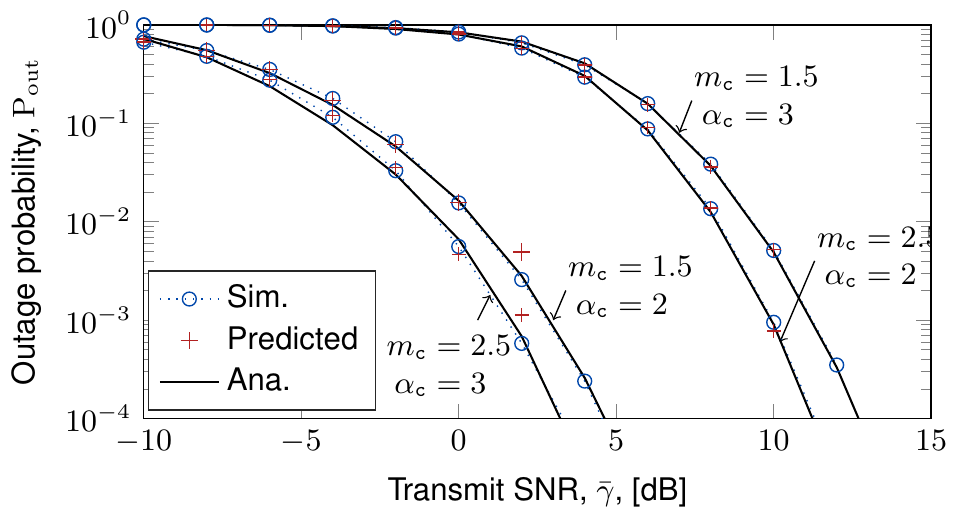}
\caption{ OP against $\bar{\gamma}$ subject to different fading conditions, with $R_\mathrm{th}=5, N=20$. It is noted that predicted results are obtained by the trained DNN associated with inputs in Table~\ref{table_parameters}, whereas analytical and simulation results are generated using fixed parameters.}
\label{fig_Pout_fading_change}
\end{figure}

As shown in Fig.~\ref{fig_Pout_fading_change}, the analytical, simulation, and predicted results are perfectly corroborated, which validates the approach in Theorem \ref{theorem_cdf_Z2} and demonstrates the accuracy of the trained DNN. It can also be observed that the outage performance is sensitive to the channel conditions, e.g., line-of-sight (LoS) strength and shadowing severity. Indeed, for a given $\bar{\gamma}$, the OP significantly decreases between the case of weak LoS, strong shadowing, i.e., $m_\mathrm{c} = 1.5, \alpha_\mathrm{c}=3$ and the case of strong LoS, weak shadowing, i.e., $m_\mathrm{c} = 2.5, \alpha_\mathrm{c}=2$.

\begin{figure}[!t]
\centering
\includegraphics[width=.9\linewidth]{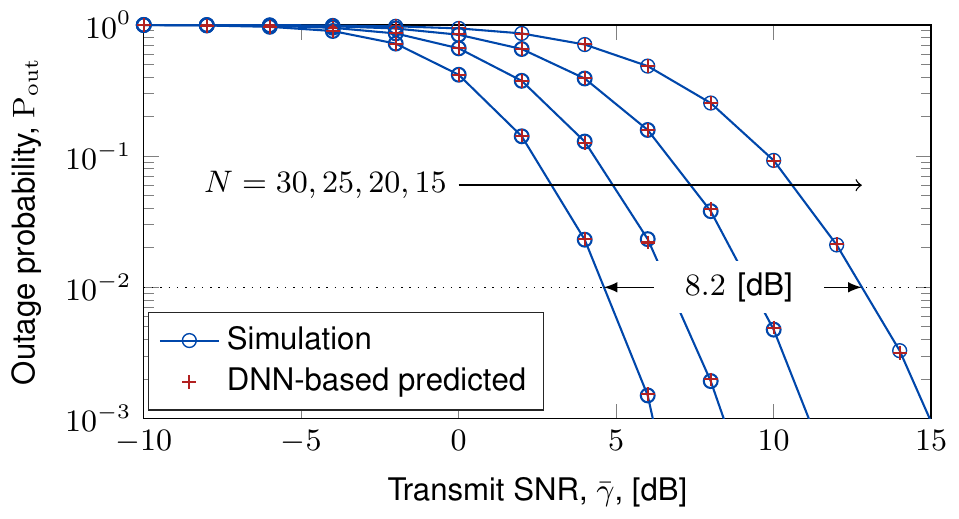}
\caption{DNN-based predicted OP against $\bar{\gamma}$ under different $N$, with $R_\mathrm{th} = 5, m_\mathrm{c}=1.5, \alpha_\mathrm{c}=3$.} 
\label{fig_Pout_N_change}
\end{figure}

In Fig.~\ref{fig_Pout_N_change}, we point out that as the number of reflecting elements increases, the power budget required to achieve a given OP is decreased, thus showing the EE of aerial-RIS. Indeed, for a given $\mathrm{P_{out}} = 10^{-2}$, $\bar{\gamma}$ drops $8.2$ dB as $N$ increases from $15$ to $30$. Again, the simulation and predicted results are well corroborated. It is noted that even if two similar feature vectors are fed to the DNN, the corresponding outputs are likely to be different because the channel gains (CSI) are RVs generated based on the feature vectors.

\begin{figure}[!t] 
	\centering
	\includegraphics[width=.9\linewidth]{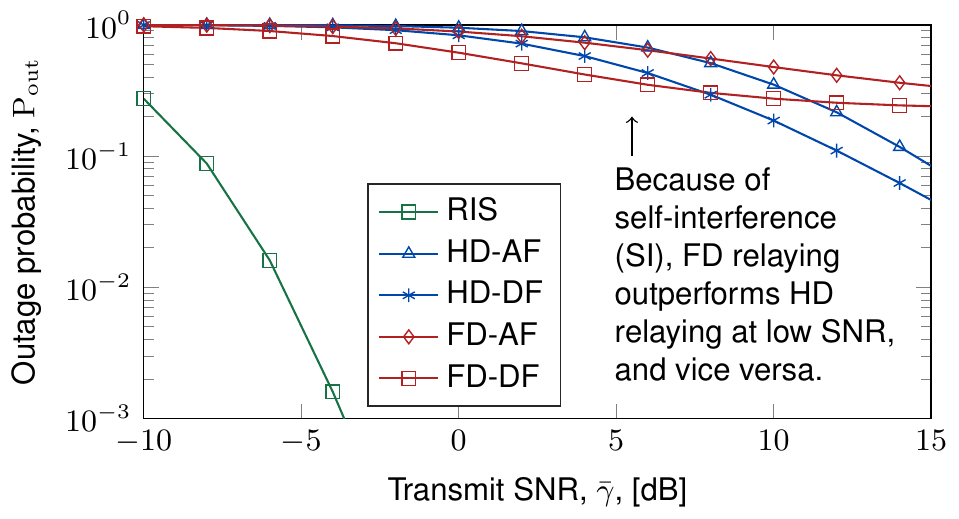}
	\caption{Performance comparison between aerial-RIS, HD-VG-AF, HD-DF, FD-AF, and FD-DF with $N = 15$, $R_\mathrm{th} = 1$, $m_\mathrm{c} =1.5$, $\alpha_\mathrm{c} = 3$.}
	\label{fig_RIS_AF_DF}
\end{figure}

In Fig.~\ref{fig_RIS_AF_DF}, we compare the performance of aerial-RIS with that of conventional time domain half-duplex (HD)-DF, HD-variable-gain (VG)-AF, full-duplex (FD)-AF, and FD-DF. For any conventional scheme, we assume that $N$ cooperative relays use maximal-ratio combining (MRC) to receive and maximal-ratio transmitting (MRT) to forward the source's signal. Here, it is observed that aerial-RIS-aided communication system significantly by far outperforms the aforementioned relaying systems.

\section{Conclusions}

In this paper, we proposed an aerial-RIS-aided wireless communication system, which takes into account practical fading channels tailored for aerial-to-ground (A2G) and ground-to-aerial (G2A) communications. We proposed a new mathematical framework to derive a tight approximate closed-form expression for the OP. Further, considering mobile environment, we developed a DNN model to deal with the 3D spatial excursion of the aerial-RIS and to predict the OP online. It was shown that (\textit{i}) the OP is significantly sensitive to the channel conditions, (\textit{ii}) the aerial-RIS provides EE, and specifically the higher the reflecting elements installed, the lower the power budget needed to achieve a given OP, and (\textit{iii}) aerial-RIS-aided system outperforms conventional dual-hop relaying systems.

\appendices
\section{Proof of Lemma \ref{lemma_matching_Gamma}} \label{apx_proof_lemma_GG_LL}
Since reflecting elements are installed closely to each other, e.g., less than a half-wave-length, individual channels in $\vec{h}_{\src \ris}$ and $\vec{h}_{\ris \des}$ are assumed to be independent and identically distributed  (i.i.d.). Thus, $\lambda_{\src r} = \lambda_\ids$, $\lambda_{r \des} = \lambda_\idd$, $\forall r$, where $\lambda \in \{m,\Omega, \alpha, \beta\}$.
However, $\Gsr$ and $\Grd$ are independent but not necessarily identically distributed (i.n.i.d.) Nakagami-$m$ RVs. Knowing that $f_{XY} (z) = \int_0^\infty \frac{1}{x} f_Y \left(\frac{z}{x}\right) f_X (x) d x$, with $X$ and $Y$ are non-negative RVs, we have
\begin{align}
f_{G_r} (z) &= \int_0^\infty (1/x) f_{\Gsr} (z/x) f_{\Grd} (x) d x \nonumber \\
&= \frac{4  ( m_\ids/ \Omega_\ids )^{m_\ids} ( m_\idd / \Omega_\idd )^{m_\idd}
 }{\Gamma(m_\ids) \Gamma(m_\idd)} z^{2m_\ids - 1} \nonumber \\
 &\quad \times \int_0^\infty \!\!\! x^{2 m_\idd - 2 m_\ids - 1} e^{-\frac{m_\ids z^2}{\Omega_\ids x^2} - \frac{m_\idd x^2}{\Omega_\idd}} d x.
\end{align}
By making use of \cite[(3.478.4)]{Gradshteyn2007}, an exact closed-form expression for the PDF of $G_r$ is attained as 
\begin{align} \label{pdf_G_r}
	f_{G_r} (z) \!=\! \frac{4 z^{m_\ids + m_\ids - 1}}{\Gamma(m_\ids) \Gamma(m_\idd)} \Upsilon_\idg^{\frac{m_\ids +m_\idd}{2}} \!\! K_{m_\idd - m_\ids} \!\! \left( 2z \sqrt{ \Upsilon_\idg } \right).
\end{align}
From \eqref{pdf_G_r}, by making use of \cite[(6.561.16)]{Gradshteyn2007} and after some mathematical manipulations, the $n$-th moment of $G_r$, $\mathbb{E}[G_r^n]$, can be attained as
\begin{align} \label{eq_mu_G_r_k}
	\mathbb{E}[G_r^n] = \Upsilon_\idg^{-\frac{n}{2}} \frac{ \Gamma(m_\ids + n/2) \Gamma(m_\idd + n/2) }{\Gamma(m_\ids) \Gamma (m_\idd)}.
\end{align}
Using the PDF of $G_r$ in \eqref{pdf_G_r} makes the derivation of the closed-expression for the PDF of $Z$ intractable. To circumvent this problem, by exploiting the statistical characteristics of $G_r$ obtained in \eqref{eq_mu_G_r_k}, we use the \textit{method of moments} (also known as the \textit{moment-matching technique}) to fit the PDF of $G_r$ to a Gamma distribution as done in Lemma \ref{lemma_matching_Gamma}. Specifically, we match the first- and second-order moments of $G_r$ to a Gamma RV $X \sim \mathrm{Gamma}(\nu_X, \zeta_X)$, i.e., $\mathbb{E}[G_r] = \mathbb{E}[X]$ and $\mathbb{E}[G_r^2] = \mathbb{E}[X^2]$. By solving this system of equations, $\nu_X$ and $\zeta_X$ can be obtained as \cite{Tahir_LWC_2021}
\begin{align} \label{eq_matched_moments}
\nu_X &= \frac{\mathbb{E}[G_r]^2}{\mathbb{E}[G_r^2] - \mathbb{E}[G_r]^2}, \quad \zeta_X = \frac{\mathbb{E}[G_r^2] - \mathbb{E}[G_r]^2}{\mathbb{E}[G_r]}.
\end{align}
Thus, we can rewrite that $\zeta_X = \mathbb{E}[G_r] / \nu_X$. Let $m_\idg \triangleq \nu_X$ and $\Omega_\idg \triangleq \mathbb{E}[G_r]$. From \eqref{eq_mu_G_r_k}, we have $\mathbb{E}[G_r^2] = \Omega_\ids \Omega_\idd$, and thus, we attain the PDF of $G_r$ in \eqref{eq_G_r_Gamma}.

We now turn our focus on $\tilde{L}_r$. It is noted that $L_\ids$ and $L_\idd$ are i.n.i.d. inverse-Gamma RVs.
Due to the fact that $f_X (x) = 2x f_{X^2} (x^2)$ and $f_{\frac{1}{X}} (x) = \frac{1}{x^2} f_X \left(\frac{1}{x}\right)$, for $x>0$, from \eqref{eq_PDF_IG}, the PDF of $1/\sqrt{L_\mathrm{c}}$ can be obtained as
\begin{align}
	f_{\frac{1}{\sqrt{L_\mathrm{c}}}} (z) = \frac{2}{\Gamma(\alpha)} \beta^\alpha z^{2 \alpha - 1} e^{-\beta z^2}, z>0.
\end{align}
Recall that $\tilde{L}_r = 1/ \sqrt{L_r}$, and $L_r = \Lsr \Lrd$, after some mathematical manipulations, the PDF of $\tilde{L}_r$ is obtained as
\begin{align}
&f_{\tilde{L}_r} (z) = \int_{0}^{\infty} \frac{1}{x} f_{\frac{1}{\sqrt{L_\idd}}} \left(\frac{z}{x}\right) f_{\frac{1}{\sqrt{L_\ids}}} (x) d x \nonumber \\
&= \frac{2 \beta_\ids^{\alpha_\ids} \beta_\idd^{\alpha_\idd} z^{2 \alpha_\idd -1} }{\Gamma(\alpha_\ids) \Gamma(\alpha_\idd)}  \!\!\! \int_0^\infty \!\!\!\!\!\! x^{2(\alpha_\ids - \alpha_\idd - 1)} e^{- \frac{\beta_\idd z^2}{x^2} - \beta_\ids x^2} d x^2.
\end{align}
Using \cite[(3.478.4)]{Gradshteyn2007} and after some derivation steps, an exact closed-form expression for the PDF of $\tilde{L}_r$ can be obtained as
\begin{align} \label{eq_n_moment_1_sqrtLr}
	f_{\tilde{L}_r} (z) = \frac{4 (\beta_\ids \beta_\idd)^{(\alpha_\ids + \alpha_\idd)/2} }{ \Gamma(\alpha_\ids) \Gamma(\alpha_\idd) } z^{\alpha_\ids + \alpha_\idd - 1} \! K_{\alpha_\ids - \alpha_\idd} \! (2z \sqrt{\beta_\ids \beta_\idd}).
\end{align}

Using \eqref{eq_n_moment_1_sqrtLr}, the $n$-th moment of $\tilde{L}_r$ can be derived as
\begin{align}
	\mathbb{E} [(\tilde{L}_r)^n] = \frac{ \Gamma(\alpha_\ids + n/2) \Gamma(\alpha_\idd + n/2) }{\Gamma(\alpha_\ids) \Gamma(\alpha_\idd)} (\beta_\ids \beta_\idd)^{-n/2}.
\end{align}
With similar steps as in \eqref{eq_matched_moments} for the case of $G_r$, let $\Omega_\idl \triangleq \mathbb{E} [\tilde{L}_r]$ and $m_\idl \triangleq \Omega_\idl^2/[\mathbb{E}[\tilde{L}_r^2] - \Omega_\idl^2]$, one can obtain the matched PDF of $\tilde{L}_r$ as in \eqref{eq_L_r_Gamma}. This completes the proof of Lemma \ref{lemma_matching_Gamma}.

\bibliographystyle{IEEEtran}
\bibliography{References_Aerial_RIS}

\end{document}